\documentstyle[12pt]{article}
\newcommand{\be} {\begin{equation}}
\newcommand{\ee} {\end{equation}}
\newcommand{\ba} {\begin{eqnarray}}
\newcommand{\ea} {\end{eqnarray}}

\begin{document}
preprint: THU-98/03 ; ULB-TH/99-05 \\

\vskip 1.5 truecm
\centerline{\Large\bf{How the Change in Horizon Area 
}}\vskip .3 truecm
\centerline{\Large\bf{Drives Black Hole Evaporation}}
\vskip 1.5 truecm
\centerline{{\bf S. Massar}}
\centerline{Service de Physique Th\'eorique, Universit\'e Libre de
Bruxelles,}
\centerline{CP 225, Bvd. du Triomphe, B1050 Brussels,  Belgium} 
\centerline{e-mail: smassar@ulb.ac.be}
\vskip 1.5 truecm
\centerline{{\bf R. Parentani}}
\centerline{Laboratoire de Math\'ematiques et Physique Th\'eorique,
CNRS UPRES A 6083,}
\centerline{Facult\' e des Sciences, 
Universit\'e de Tours, 37200 Tours, France. }
\centerline{e-mail: parenta@celfi.phys.univ-tours.fr}
\vskip 1.5 truecm

\centerline{\bf{Abstract}}

We rephrase the derivation of black hole radiation
so as to take into account, at the level of transition amplitudes,
the change of the geometry induced by the emission process.
This enlarged
description reveals that the dynamical variables
which govern the emission 
are the horizon area and its conjugate time variable.
Their conjugation is established through the boundary term at 
the horizon which must be added to the canonical action of general 
relativity in order to obtain a well defined action principle when 
the area varies. 
These coordinates have already been used  
by Teitelboim and collaborators 
to compute the partition function of a black hole.
We use them to show 
that the probability to emit a particle is
given by $e^{- \Delta A/4}$ where $\Delta A$ is the decrease in 
horizon area
induced by the emission. 
This expression improves Hawking
result which is governed by a temperature (given by the surface gravity)
in that the specific heat of the black hole is no longer neglected.
The present derivation of quantum black hole radiation is based 
on the same principles which are used to derive the first law of 
classical black hole thermodynamics. Moreover 
it also applies to quantum processes associated with
cosmological or acceleration horizons. 
These two results indicate that
not only black holes but all event horizons possess 
an entropy which governs processes according to 
quantum statistical thermodynamics.

\newpage
\section{Introduction}

There are two possible approaches to the 
gravitational back reaction to Hawking radiation. The first is to
develop a microscopic theory of quantum gravity and to use it to
calculate the properties of black holes. This program has been
partially realized in the context of super-string theory\cite{calmad}. 
The second is to use Hawking's calculation\cite{H} as a starting point 
to compute the gravitational corrections to black hole evaporation.
Hopefully these two approaches  should meet in some middle ground.

In the 
second approach, the back reaction has been
addressed along two complementary lines. The first is the
``semi-classical'' theory wherein one first computes
the mean (quantum average) energy-momentum tensor
of the quantized fields propagating on the background geometry
and then solves the Einstein equations driven by 
this mean value.
The solution describes an evaporating geometry
characterized by the shrinking of 
the horizon area\cite{Bard, PP, massar}.
In this treatment, the metric remains classical 
and unaffected by the quantum fluctuations 
of the energy-momentum tensor.

The second line of attack is to take into account the dynamics
of gravity at the level of transition amplitudes before performing
the quantum average over the configurations of the radiation field. 
Since Hawking radiation is derived from quantum field theory, the most
natural procedure
would be to use Feynman rules.
However, the ill-defined ultraviolet behavior of quantum
gravity has so far prevented this approach  from being successfully
followed. Thus, one has to resort to less ambitious schemes. 

Some interesting insights have been obtained by taking 
the gravitational back reaction into account at the classical level,
before computing transition amplitudes. To this end,
the matter action $S_{matter}$ in a given geometry 
is replaced by the action of matter plus gravity $S_{matter+gravity}$. 
A concrete model has been 
developed by Keski-Vakkuri, 
Kraus and Wilczek (KKW) \cite{KW1,KW2,KK}.
It describes the propagation of a massless spherically symmetric
self-gravitating shell in a black hole geometry.
Having computed the new action, 
they postulate that the wave function of a 
shell is given by the WKB form $e^{iS_{m+g}}$. 
Using these waves, they derive the black hole emission
amplitudes of uninteracting (dilute gas approximation) shells.
A similar approach
has been used in a Euclidean framework in \cite{MP}
following the techniques developed in \cite{HHR}.
In this case, the
probability for the black hole to emit a particle is expressed in terms of the
action $S_{m+g}$ of a self gravitating instanton.

The striking result of these works is that the probability for a
black hole of mass $M$ to emit a particle of energy $\lambda$ is given by 
\begin{equation}
P_{M \to M - \lambda} = N(\lambda, M ) e^{- \Delta A (\lambda, M) /4 }
\label{R}\end{equation}
where $\Delta A (\lambda, M)= A(M) - A(M -\lambda)$
is the decrease of the area of the black
hole horizon.
$N$ is a phase space 
(also called grey body) factor which cannot be calculated 
in these approximation schemes. 

Eq. (\ref{R}) replaces Hawking's result
$P_{M \to M - \lambda} = N e^{- 2 \pi \lambda / \kappa}$
which is characterized by a temperature $T_H = \kappa / 2 \pi$ 
defined by the surface gravity $\kappa$.
To first order in $\lambda$, the first law\cite{WALDBOOK} of
classical black hole thermodynamics $d E = {\kappa \over 8 \pi } dA$
guarantees that the two expressions for $P_{M \to M - \lambda}$
coincide.
This suggests
that Hawking radiation and the first law both stem from the same 
principle.
This is far from obvious since the first law relates neighboring 
classical solutions of general relativity
whereas black hole radiance is derived from QFT in a given geometry.

The aim of the present work is to reveal their common origin
and to establish the universal validity of eq. (\ref{R}).
We shall show that both follow from
the use of the complete action $S_{m+g}$ in place of the 
matter action in a given geometry.
The reason is that the emission of a matter quantum can now be
viewed as the transition between two neighboring black hole states.
In this transition, the energy is transfered from
the hole to the radiation, 
a feature absent in the 
derivation of black hole radiance based on Bogoliubov coefficients
evaluated in a given geometry. 
Moreover, the relevant dynamical quantity governing this transition is  
the difference of actions: $S_{m+g} (final) - S_{m+g} (initial)$.
Then, as in deriving the first law\cite{CT}, this difference can be
reexpressed as a difference of boundary terms at the horizon. From 
this, it is easy to show that the dynamical quantity which governs 
emission rates is equal to the change in horizon area divided by 4.

To demonstrate this result,
we make use of the appropriate coordinate system to
describe processes inducing changes in area. These are the boost
parameter $\Theta$ and its conjugate variable, the area of the horizon
$A/8\pi$. These variables where used by Teitelboim and collaborators
\cite{CT,Tp,T,CT2,BTZ} to compute the partition function of the black hole
{\it \`a la} Gibbons-Hawking\cite{GH} starting from the appropriate
classical action. In this case, they showed that one must add
to the canonical action ($pdq - H dt$) a boundary term equal 
to $\Theta A/8 \pi$. 
In our context, it is through this boundary term that we shall obtain 
eq. (\ref{R})
and establish the relationship with the first law.

We have organized this paper as follows. After a brief review
of Hawking's derivation formulated in a fixed geometry,
we introduce the reader to the variables 
$\Theta$ and $A$ and to their role in boundary terms at the horizon.
In particular we emphasize that the particular form of the boundary term
is dictated by the physical process considered. For the emission of
particles by a black hole, the appropriate action is the 
canonical action $S_{m+g}$ supplemented by $\Theta A /8 \pi$.

This action is used in section \ref{SProb} to compute the transition rates
of a detector at a fixed radius of the black hole. The usefulness of
introducing a detector is that the transition amplitudes are expressed
in terms of the overlap of the initial and final states of the
detector+black hole complex. Both are stationary 
eigenvectors of $i\partial_\Theta$ with eigenvalue equal to the area
of the black hole. In this setting the ADM mass is fixed and therefore
the time at infinity cannot be used to parameterize the
evolution\cite{PW}. Instead one must use $\Theta$ time. It is then 
straightforward to 
show that the ratio of transition rates of the detector is
given by $e^{-\Delta A/4}$.

Even though the introduction of a detector is useful, it is
not intrinsic to black hole radiation. This is why in section
\ref{secKKW} 
we reconsider the KKW model which only makes appeal to the action of
the particles emitted by the black hole. By using $\Theta$ and $A$ we
shall recover KKW's result in very simple terms and make contact with
the former derivation.
We shall also show that eq. (\ref{R})
follows directly from the {\it universal} form of out-going trajectories
in the near horizon geometry and the specification 
that the field configurations be in (Unruh) vacuum.
%
Thus eq. (\ref{R}) applies to all emission processes in the presence
of horizons, including charged and rotating black hole, cosmological
and acceleration horizons (in the latter case the horizon area is infinite, 
but differences are finite and well-defined, see \cite{HHR,MP,suh2}). 

The universal validity of this derivation is the main result of this
paper. It proves that the area of all event horizons determines 
the gravitational statistical entropy 
{\it available}\cite{PP2} to these quantum processes.
In this we give statistical foundations\cite{SORKIN2}
to the relation between horizon thermodynamics and 
Einstein's equations exhibited by Jacobson\cite{jac}.

\section{Hawking Radiation}\label{two}

In this section we fix the notations and review two standard
derivations of black hole radiation. 
In the first derivation, following Unruh\cite{U}, we 
introduce a two level atom coupled to the radiation field
and whose position is fixed.
Then, using Einstein's argument, one determines the 
distribution of massless quanta from its transition rates.
In this way one only uses basic quantum mechanical rules.
The second approach is more intrinsic and closer to Hawking's 
derivation\cite{H}. Black hole radiation is established 
through the Bogoljubov transformation relating in-modes which
determine the state of the radiation field before the collapse
(which shall be taken for simplicity to be in-vacuum)
 and out-modes which define the particles emerging 
from the hole and found at infinity.

The metric of a Schwarzschild black hole is
\begin{equation}
ds^2 = -(1 - {2 M \over r}) dt^2  +(1 - {2 M \over r})^{-1} dr^2 + r^2
d\Omega^2 \quad .
\end{equation}
We introduce the light like coordinates $u$ and $v$:
\begin{equation}
v,u = t \pm r^*\quad , \quad r^* = r +  2M \ln \vert r- 2M \vert
\end{equation}
and the Kruskal coordinates $U_K$ and $V_K$:
\begin{equation}
U_K = - 
{ 1 \over \kappa} e^{-\kappa u} \quad , \quad 
V_K= { 1 \over \kappa} e^{\kappa v} 
\end{equation}
where $\kappa= 1/4M$ is the surface gravity.

For a black hole formed by the collapse of a 
spherically symmetric star, the outgoing modes,
solutions of the Dalembertian equation
have the following form near the horizon,
i.e. for $r - 2M \ll 2M$:
\begin{equation}
\phi_{\omega l m} = { e^{-i\omega U_K} \over \sqrt{4 \pi \omega}}
{Y_{lm}(\Omega) \over r} \label{Umodes}
\quad .
\end{equation}
Further away from the horizon this expression 
is no longer exact because of the potential barrier which 
surrounds the black hole. For simplicity, throughout the article,
we shall neglect the transmission coefficients (also called grey body factors)
 induced by this barrier since they cancel out 
from the ratio of the transition rates 
which determines the equilibrium distribution around an eternal black
hole, see eq. (\ref{Prob}) below. 
This cancellation also applies to charged and rotating black holes 
when one takes
into account the ``work term'' exerted by the electric potential or the 
angular momentum of the hole, see \cite{Gibbons, BD}. 

The modes eq. (\ref{Umodes}) are associated with the 
out-going particles as seen by infalling observers. 
Indeed, for $r- 2M\ll 2M$, the proper time lapse of infalling observers 
is proportional to $\Delta U_K$. This is simply seen by 
re-expressing the Schwarzschild metric in $U_K, V_K$ coordinates:
 $ds^2 \simeq - dU_KdV_K + r^2 d^2 \Omega$. Hence near the horizon,
 for infalling observers, the modes eq. (\ref{Umodes}) have positive
 frequency.

Using these modes, the field operator can be decomposed as:
\begin{equation} 
\Phi = \sum_{\omega,l,m} a_{\omega l m}
\phi_{\omega l m}  + {\rm h.c.} + {\rm ingoing\ modes} 
\quad .
\end{equation}
By definition, the in-vacuum state, denoted $|0_U\rangle$, 
is annihilated by the $a_{\omega l m}$ operators. 
In the literature it is often called the Unruh vacuum.
The fact that it is annihilated by all $a_{\omega l m}$ 
guarantees that infalling observers 
experience vacuum conditions as they cross the future horizon.

Consider now a particle detector at fixed radius $R$ of the black hole
and angular position $\Omega$. It has two levels $|e \rangle$ 
and $|g \rangle$ of
energy $E_e$ and $E_g$ with $\Delta E = E_e - E_g >0$.
For simplicity we shall take $R$ to  be very large
($R \gg 2M$).  
Then $t$
is the proper time of the detector and 
$ \Delta E$ is the energy gap as measured from
$r = \infty$. For smaller $r$,
one should take into account the gravitational red shift.

In the interaction representation,
the coupling of the detector to the field $\Phi$ is given by
\begin{equation}
H_{int} = \gamma
\Phi(t,R,\Omega)\left( e^{+i \Delta E t} |e \rangle \langle g| + {\rm
h.c.} 
\right)
\label{hinter}
\end{equation} 
where $\gamma$ is the coupling constant.

When the detector is initially 
in its ground state, the state of the system 
(detector plus radiation field $\Phi$) is $|0_U \rangle|g \rangle$. 
In the interacting picture, at late times and to first
order in $\gamma$, this state becomes:
\begin{eqnarray}
|\psi_g \rangle&=&|0_U \rangle|g \rangle - i \gamma \int dt 
e^{i \Delta E t} \Phi(R,\Omega,t)
|0_U \rangle|e \rangle\nonumber\\
&=&|0_U \rangle|g \rangle - i \gamma \sum_{\omega l m}\int dt 
e^{i \Delta E t} e^{-i
\omega  C e^{-\kappa t} } {Y_{lm}^*(\Omega)\over R \sqrt{4 \pi \omega}}
a^\dagger_{\omega l m}
|0_U \rangle|e \rangle 
\nonumber\\
&=&|0_U \rangle|g \rangle + \sum_{\omega l m}
{B}_{g \to e, \omega,l,m}
a^\dagger_{\omega l m}|0_U \rangle|e \rangle 
\label{gtoe}
\end{eqnarray}
where $C= (1/\kappa) \exp (\kappa R + {1 \over 2}\ln (R - 2M))$.

Similarly if the detector was initially in its excited state, the
final state would have been
\ba
|\psi_e \rangle &=&
|0_U \rangle|e \rangle - i 
\gamma \sum_{\omega l m}\int dt e^{-i \Delta E t} e^{-i
\omega C e^{-\kappa t}}  {Y_{lm}^*(\Omega)\over R \sqrt{4 \pi \omega}}
a^\dagger_{\omega l m}
|0_U \rangle|g \rangle
\nonumber\\
&=& |0_U \rangle|e \rangle + \sum_{\omega l m}
{B}_{e \to g, \omega,l,m}a^\dagger_{\omega l m} |0_U \rangle|g \rangle 
\quad .
\label{etog}
\ea 

Since we are interested in determining the population 
of quanta seen by the detector, we only need to 
compute the ratio of the transition rates.
By replacing $t$ by $t + i \pi/ \kappa $ in the integral 
governing the transition amplitude ${B}_{e \to g, \omega,l,m}$, 
one obtains 
$B_{g \to e} = B_{e \to g}^* e^{- \pi \Delta E /\kappa}$ for all 
$l,m,\omega$, 
see \cite{PB, GO}.
Thus the ratio of the transition  probabilities is:
\begin{equation}
{|{B}_{g \to e, \omega,l,m}|^2 \over |{B}_{e \to g, \omega,l,m}|^2}
=e^{- {2\pi \over \kappa} \Delta E} \label{Prob}
\quad .
\end{equation}
This corresponds to the rates in a thermal bath at temperature $
\kappa / 2 \pi = 1/8 \pi M$. 
Then, using Einstein's argument, one 
obtains that the quanta of the radiation field are distributed according to
the Planck distribution.

This derivation is equivalent to calculating the 
Bogoljubov transformation between Unruh modes, eq. (\ref{Umodes}),
 and the out-modes defined below in eq. (\ref{outmodes}). 
Indeed the 
transition amplitudes ${B}_{g \to e, \omega,l=0}$ and
${B}_{e \to g, \omega,l=0}$ are proportional to 
the Bogoliubov coefficients
$\alpha_{\omega \lambda} $ and $\beta_{\omega \lambda}$ 
when $\Delta E = \lambda$, see \cite{U, PB}.
This relationship provides a physical interpretation of 
Bogoljubov coefficients as transition amplitudes. 
A concept that we shall rediscuss below.

We now turn to the second derivation in which 
the spectrum of emitted particles is determined only
in terms of solutions of the Dalembertian 
with a given frequency $i\partial_t=\lambda$. There are
two positive norm modes for each value of $\lambda$ which identically 
vanish either inside or outside the horizon:
\begin{eqnarray}
\varphi_{\lambda,+} &=& {e^{- i\lambda u}\over \sqrt{4 \pi \lambda}} 
{ 
 1 \over r}
 \theta (r-2M) \quad , \quad \lambda > 0 \quad ,\nonumber\\
\varphi_{\lambda,-} &=& {e^{+ i\lambda u}\over \sqrt{4 \pi \lambda}} 
{  
 1 \over r}
 \theta (2M-r) \quad , \quad \lambda > 0 \quad .
\label{outmodes}
\end{eqnarray}
Once more we have neglected  the potential barrier outside 
the black hole and for simplicity we have considered only s-waves
($l=0$).

The modes $\varphi_+$ define the out-quanta, i.e. 
those used by a static observer around the black hole
to describe the presence or absence of out-going particles. 
We can again decompose the field operator into $\varphi_\pm$ modes:
\begin{equation} \Phi = \sum_{\lambda,\pm} a_{out,\lambda \pm}
\varphi_{\lambda, \pm}  + {\rm h.c.} + {\rm ingoing\ modes} 
\quad .
\end{equation}
By definition the $a_{out,\lambda \pm}$ annihilate the out vacuum, 
$|0_{out}\rangle$.

It is now appropriate to introduce a third set of modes $\phi_{\lambda, \pm}$
which possess the following properties.
They are eigenmodes 
of $i\partial_t=\lambda$ and are composed only 
of positive frequency modes, eq. (\ref{Umodes}),
which define Unruh vacuum.
The simplest way to implement this last condition is 
to express the out-modes in Kruskal coordinates:
$\varphi_{\lambda \pm} \simeq (\mp U_K)^{\pm i \lambda / \kappa}
\theta(\mp U_K)$. Since $\Delta U_K$ is
proportional to the proper time of an infalling observer,
$\phi_\lambda$ must be the linear combination of $\varphi_{\lambda
\pm}$ which is analytic and bounded in the 
lower half of the complex $U_K$ plane. Upon requiring also that
the $\phi_{\lambda \pm}$  have unit
norm, one obtains
\begin{eqnarray}
\phi_{\lambda +} = {1 \over \sqrt{ 1 - e^{- 2 \pi \lambda/ \kappa}}}
\left( \varphi_{\lambda +} + e^{-  \pi \lambda /\kappa}
\varphi_{\lambda -}^* 
\right )\quad , \quad \lambda > 0 \quad , \nonumber\\
\phi_{\lambda -} = {1 \over \sqrt{ 1 - e^{- 2 \pi \lambda/ \kappa}}}
\left( \varphi_{\lambda -} + e^{-  \pi \lambda /\kappa}
\varphi_{\lambda +}^* 
\right ) \quad , \quad \lambda > 0 \quad .
\label{US}\end{eqnarray}
One  can verify by evaluating the overlap of 
$\phi_\lambda$ and $\phi_\omega$ that the $\phi_\lambda$ are linear
combinations of the $\phi_\omega$ with no $\phi_\omega^*$ component, 
see e.g. \cite{GO}.
One can then decompose the field operator in terms of these new modes:
\begin{equation} \Phi = \sum_{\lambda,\pm} a_{\lambda \pm}
\phi_{\lambda, \pm}  + {\rm h.c.} + {\rm ingoing\ modes}
\quad . 
\end{equation}
Then, Unruh vacuum is annihilated by the $a_{\lambda \pm}$ operators.

The weights in
eq. (\ref{US}) define  the Bogoljubov coefficients $\alpha_\lambda$
and $\beta_\lambda$.
Their ratio satisfies
\begin{equation}
{ | \beta_ \lambda |^2 \over | \alpha_ \lambda |^2}
= e^{- 2 \pi  \lambda/\kappa } \quad .
\label{eqex}
\end{equation}
Since $ | \beta_ \lambda |^2$ determines the mean number of out quanta
of energy $\lambda$ in Unruh vacuum, 
eq. (\ref{eqex}) implies that Unruh
vacuum is a thermal distribution of out-particles 
at temperature $\kappa /2  \pi$, in agreement with eq. (\ref{Prob}).

We will find it convenient below to use the same argument, but
rephrased in coordinate systems which are regular on the 
future horizon and which lead to a static metric.  This second condition
implies that the time  parameter is
proportional to $t$ at fixed $r$.
An example is given by the Eddington-Finkelstein coordinates
$v,r,\Omega$ in which the metric has the form:
\begin{equation}
ds^2 = -(1-2M/r) dv^2
+ 2dvdr + r^2 d\Omega^2 \quad .
\end{equation}
Near the horizon the metric takes the simple form
$ds^2 \simeq 2dvdr + r^2 d\Omega^2$ which shows that $v,r$ are
inertial coordinates. 
Moreover since $-dr$ is proportional to $dU_K$,
the momentum $p_r$ plays the role of the 
frequency $\omega$ of eq. (\ref{Umodes}) 
and
one can translate the analytical condition implementing Unruh vacuum
in terms of $r$: to obtain 
a $\phi$ mode, one analytically continues $\varphi_+$ in
the upper half complex $r$ plane to define its value for $r < 2M$.
As shown in \cite{DamourRuff}, one immediately obtains eq. (\ref{US}).
In Section 5, it is through the analytical behavior in $r$
of the modified modes that we shall determine the corrections to 
eq. (\ref{eqex}).

In conclusion of this Section, 
we wish to emphasize the following point.
In the transitions described in eq.  (\ref{gtoe}),
there is a transfer of energy from the radiation
field to the detector but the black hole
mass stays constant. Similarly, 
upon computing the Bogoliubov coefficients in eq. (\ref{US}),
the geometry is unaffected. 
In these descriptions of black hole radiation,
there is no transfer of energy from the hole
into radiation. The notion of black hole evaporation
only arises when the mean energy momentum tensor 
of the radiated quanta
is put on the r.h.s of Einstein's equations wherein it drives
the shrinking of the horizon area\cite{H}-\cite{massar}. 

This artificial two step procedure results
from the original sin: to have decided to work in a given 
geometry. As we shall see in Sections 4 and 5,
upon working with a recoiling geometry, energy 
conservation will be taken into account at the 
level of amplitudes, as in the Compton effect.
Then, the black hole will act as a conventional reservoir of energy:
when delivering heat to the external world it loses 
the corresponding energy.

\section{Boundary terms in the Einstein-Hilbert Action}\label{boundary}

In this section, following \cite{CT, Tp, T},
 we introduce coordinates which have an intrinsic
geometric interpretation near the horizon. These coordinates are the
hyperbolic angle $\Theta$, the transverse coordinates along the
horizon $x^i_\perp$ and the radial proper distance from the
horizon $\rho$. In terms of these coordinates, the metric near the 
horizon takes the universal form
\begin{equation}
ds^2 \simeq -\rho^2 d\Theta^2 + d\rho^2 + \gamma_{ij}dx^i_\perp dx^j_\perp
\quad .
\label{dsH}
\end{equation}
For simplicity we have written the metric for a spherically symmetric 
horizon, for the general case we refer to \cite{T}.
The area of the horizon is $A =\int dx^2_\perp \sqrt{\gamma}\vert_{\rho =0}$.
In the case of a Schwarzschild on-shell solution,
$\Theta = \kappa t_\infty$ and $\rho
\simeq \sqrt{ 8M(r-2M)}$ where $t_\infty$ 
is the proper time at spatial infinity.

The universal form of the metric in these coordinates
implies that if we use them to describe 
a process near a particular horizon, 
the description of the same process in the vicinity of any
other horizon will be identical. 
The developments presented in sections \ref{SProb} and \ref{secKKW}
for a Schwarzschild black hole thus also apply
to charged and rotating holes\cite{CT2} and to 
cosmological and acceleration horizons.
Furthermore because these coordinates lead to such 
a simple form for the metric, 
physical process occurring near the horizon
 will be most simply described in these coordinates.
In this section, this will be illustrated by
considering the boundary terms at the horizon that arise 
in the Hamiltonian formulation of general relativity. 
In the next sections we shall see that these coordinates
are also well adapted to describe particle production near event
horizons.

We start the analysis of the boundary terms in the canonical action 
for matter and gravity
\begin{equation}
 I_{can}= \int^t_0 dt' \left\{
\pi^{ab} \dot g_{ab} + p \dot q - N H - N^i H_i
\right\}
 \label{S}
\end{equation}
where $g_{ab}$ is the spatial metric, $\pi^{ab}$ its conjugate
momentum, $q$ and $p$ the coordinates and momentum of matter, $N$ and
$N^i$ the lapse and shift, and $H$ and $H_i$ the energy and momentum
constraints. 
In what follows, we focus on metrics $g_{ab}$ 
which have two spatial boundaries. 
We suppose that at one boundary the metric is
asymptotically flat, that is $g_{ab}$ and $\pi^{ab}$ tend to
their value in flat space as the proper distance $\rho \simeq r$ 
tends to infinity, see \cite{RT} for the precise behavior 
one imposes. We also suppose that 
at the other boundary the metric can be put in the form
eq. (\ref{dsH}), see \cite{T} for the precise conditions imposed 
at this boundary. Thus we are considering 
the class of metrics which, on shell and in vacuum, will correspond 
to one of the asymptotically flat quadrants of an eternal black hole.

Because these metrics have spatial boundaries, 
it may be necessary to 
add boundary terms to $ I_{can}$ in order for it to be  
stationary on the solutions of the equations of motion.
We now review these features, starting with the boundary 
at infinity. Details of the calculations will not be presented. 
They can be found in many papers, 
see for instance \cite{RT}\cite{K}\cite{BTZ}\cite{T}. 
Let us first emphasize that the following considerations
are superfluous for classical physics, i.e. for the determination
of the solutions of the equations of motion. However,
when considering quantum kernels, transition amplitudes
or partition functions, they cannot be ignored since
 the WKB value of the quantum phase is determined 
by the action.

To identify the boundary terms, following 
\cite{Tp}, we compute the total variation of the
action given in eq. (\ref{S}).
By total variation we mean that we are considering the linear change 
of $I_{can}$ due to an arbitrary change of all its variables. 
We note that this variation  can be performed around any
configuration, i.e., on-shell or off-shell.
We nevertheless impose that the space time 
is asymptotically flat. In this case, the first order change of $I_{can}$
contains three types of terms. First we have the contribution of the bulk, 
the 4-geometry interpolating the initial to the final 3-geometry. As usual,
this contribution vanishes when the reference configuration is on-shell.
Secondly, one has the contributions due to the change of
 the initial and final configuration.
These determine (as usual) the initial and final momenta of gravity.
Finally, one has an additional contribution arising from spatial infinity.
On shell, it is is equal to
\begin{equation}
\delta  I_{can}^{r=\infty}= t_\infty \delta M_{ADM}
\label{deltaSHam}
\end{equation}
where $t_\infty$ is the (coordinate invariant) proper time 
at spatial infinity. It is related to the
lapse function $N(r)$ by $t_\infty
 = \int_0^t dt' N (t', r=\infty)$.
$\delta M_{ADM}$ is the change in mass at infinity 
which is defined by  $\delta g_{rr}$ for large $r$.
Eq. (\ref{deltaSHam}) shows that $I_{can}$ is extremal on shell, 
i.e. its variation reduces on shell to 
$\pi^{ab} \delta g_{ab}\vert_{initial}^{final}$,
 only if $\delta  M_{ADM}=0$,
i.e. only if one varies among the sub-class of 
metrics for which $M_{ADM}$ is kept fixed. 

If one wishes to consider the Legendre conjugate ensemble
in which the asymptotic  proper time is fixed, one
must work with the action $S' =  I_{can} - t_\infty M_{ADM}$
Indeed, the variation of this new action yields the following 
asymptotic contribution
\begin{equation}
\delta S'_{r=\infty} = -  \delta t_\infty M_{ADM} \, .
\label{deltaS'} \quad 
\end{equation} 
This term vanishes when one works
 among the sub-class of metrics for which $t_\infty$ is kept fixed 
but $\delta M_{ADM}$ is arbitrary.

We now turn to the boundary term at the horizon. The analysis
 proceeds in parallel with the preceding one. 
The form of the boundary  term is dictated by the fact that one requires
 that, near the horizon, the lapse and shift vanish (or more precisely
that at $\rho =0$, the momentum $\pi^\rho_\rho$ and the derivative of
the area of the surfaces of constant $\rho$, $\partial_\rho A$, 
vanish, see \cite{T})
and hence the metric can be put in the form eq. (\ref{dsH}).
Upon varying eq. (\ref{S}) one finds a boundary term at the horizon
\begin{equation}
\delta  I_{can}^{inner \, boundary}= 
- \Theta \delta A/8 \pi \ .
\label{deltaSHam2}
\end{equation}
Here $\delta A$ is the change in horizon area and
$\Theta$
is the hyperbolic angle defined 
in eq. (\ref{dsH}). It is equal to the limit $r \to r_{horizon}$ of
$  N^2(r) t / 2 (r- r_{hor})$.
For simplicity of writing, we have again considered
only spherically symmetric 3-geometries.
For the general case, see \cite{T}.
Notice that $\Theta$ is a coordinate invariant
quantity, exactly like $t_\infty$ in the asymptotic contribution.
Moreover, as pointed out in \cite{CT}, $A/8 \pi$ and $\Theta$ 
are conjugate variables exactly like $M_{ADM}$ and $t_\infty$.
Eq. (\ref{deltaSHam2}) shows that $ I_{can}$ is extremal on-shell
 for the sub-class of metrics which have fixed  horizon area.

When one wishes to work with a fixed opening angle $\Theta$,
one must work with the action $S= I_{can}+ \Theta A /8 \pi $
since its variation yields
\begin{equation}
\delta S^{inner \, boundary} = A \delta \Theta / 8 \pi \, .
\label{dddd}
\end{equation}
$ S$ is thus extremal on-shell 
for the class of metrics which have fixed $\Theta$.
This action must be used when working in 
the Euclidean continuation of the black hole\cite{BTZ}. 
Indeed regularity of the Euclidean manifold at the horizon 
imposes a fixed Euclidean angle
given by $\Theta_E = 2 \pi$. 
The surface term in the action then contributes a
term $A/4$ to the partition function which is 
interpreted as the entropy of the
black hole. One of the main advantage of introducing 
this boundary term is to clarify the derivation of this 
 partition function which was first considered by Gibbons and
Hawking\cite{GH}.

In brief, upon considering dynamical processes 
occurring around a black hole,
there are a priori 4 actions which can
be considered according to which quantities are fixed in the 
variational principle (or in the path integral)
and therefore according to the surface terms.
Two however are rather unphysical. Indeed fixing both the ADM mass
and the horizon area is too constraining as can be seen by considering  
the vacuum spherically symmetric solutions for which  
fixing the ADM mass determines the horizon area
(in a path integral,
the kernel would vanish). For non empty 
geometries, this double specification would impose 
an unusual non-local constraint on the matter energy repartition. 
Similarly fixing both the time at infinity and the hyperbolic
angle $\Theta$ is also too constraining. Thus one is left with two
possibilities: fixing $\Theta$ and $M_{ADM}$ 
or fixing $t_{\infty}$ and $A$. 
A more mathematical reason why these are the only 
two possibilities (in the absence of matter)
is that the constraints $H=0$ and $H_i=0$ 
viewed as differential equations  
need boundary conditions in order to yield a unique solution
and fixing the ADM mass 
or the horizon area but not both provides the required 
boundary data\cite{T}.

The choice among the two remaining possibilities is dictated by physical
considerations. If $M_{ADM}$ is not fixed but $A$ is, 
this means that one is considering configurations 
in which the ADM mass is determined by
the repartition of matter surrounding the black hole while
leaving the black hole unchanged. On
the other hand, if one fixes $M_{ADM}$ while letting $A$ to be determined,
one is exploring configurations in which energy can be redistributed
between the black hole and the
surrounding matter, but with no change of the ADM mass
\footnote{
To be complete, we should perhaps point out the following
difficulty. The horizon  is
the 2D boundary common to all 3-surfaces including 
the initial and the final ones.
Hence, its area cannot vary in time.
Therefore in order to allow $A$ to vary,
one must consider a kind of regularized version 
in which the boundary of the 3-surfaces is arbitrarily close to the
horizon, but does not coincide with it. To our knowledge,
the precise procedure has yet not been completely worked out. 
We mention here the recent works\cite{W, C, A} in which 
generalized definitions of horizon 
have been proposed, mainly to allow for evolution. 
However, in what follows, we shall not need an action governing
continuous changes in area. 
Only states corresponding to constant areas will be used.}.

Clearly black hole evaporation requires the second
situation if one wants to analyze what happens
at finite $r$. 
Therefore in the next section,
we shall fix the mass at 
infinity and follow the evolution  in terms of $\Theta$
and $A$. The relevant action in this case is 
\begin{equation}
S = \int_0^t dt' \left\{ \pi^{ab} \dot g_{ab} + p \dot q - N H - N^i H_i 
\right\}
+ \Theta(t) \, A / 8 \pi
\quad . \label{S2}
\end{equation}

In 
section \ref{secKKW} we shall also consider the region
behind the horizon. Thus the formalism developed here will 
not be directly applicable. Nevertheless,  
it will still be convenient to keep the ADM mass fixed and to describe
the process in terms of $\Theta$ and $A$.

\section{Probability for detector transitions in terms of horizon area
change}\label{SProb}

In this section and the following one we calculate the transition rates
governing black hole radiation 
when the matter action in a fixed 
geometry is replaced by the sum of matter 
and Einstein-Hilbert actions.
In this new description, the 
change of the geometry induced by the transition
process is taken into account through 
the extremisation of the total action.
We start with the description of black hole radiation based
on the transitions of a static detector. 

Our aim is to show that it is the change in area
associated with the quantum jump of the detector 
which determines the transition amplitudes.
The detector is assumed to be at $r=R$ in one of the
asymptotically flat quadrants of an eternal black hole. We believe
however that our analysis also applies to black
holes formed by collapse  and that the mathematical
analysis in the two cases should be identical, see however
the last footnote.
As in Section 2, we compute the transition amplitudes in perturbation
theory to first order in  the coupling $\gamma$ 
of the detector to the field, see eq. (\ref{hinter}).
As before they are given by the overlap of the three free 
waves. Here these are the
radiation wave function and the two 
stationary states of the detector + black hole complex.
These two stationary states have two interesting properties. 
Firstly, their eigenvalue is the black hole area rather than the 
detector's energy. Secondly, since the ADM mass is fixed,
each eigenvalue is entangled 
to the corresponding detector state.
Thus, in the new description, the transitions of the
detector lead to quantum jumps from one horizon
area to the other without smooth (classical) interpolation
from one stationary geometry to the other.

We first compute the $\Theta$-time dependence of the 
wave functions associated with the two states of the 
detector when their energy is taken into
account in the definition of the background geometry. 
This amounts to evaluate twice the on-shell action, eq. (\ref{S2}).
Since the detector is at $r=const$, both classical geometries
are static. Thus the $p \dot q$ and $\pi^{ab}\dot g_{ab}$
terms in the action vanish. 
Moreover, on-shell, the constraints also vanish.
Hence, at fixed $M_{ADM}$,
 the only term contributing to $S$ is the surface term at the
horizon. This term is equal to $\Theta A_g/8 \pi $ or $\Theta A_e/8 \pi$
where $A_g$ ($A_e$) is the horizon area when the detector is in the
ground (excited) state. 
Thus the time dependence of the free (i.e. $\gamma =0$)
wave functions
are
\begin{eqnarray}
\Psi_{BH+g}(\Theta) &=& e^{i \Theta
A_g/8 \pi} \, \Psi_{BH+g}(0) \quad , \nonumber\\
\Psi_{BH+e}(\Theta) &=& e^{i \Theta
A_e/8 \pi} \, \Psi_{BH+e}(0)\quad .
\label{PsiBHd}
\end{eqnarray}
This is certainly correct in a WKB approximation.
Moreover, since in the 
absence of interactions with the radiation field
 one deals with stationary area eigenstates, the exponential form is exact. 
Of course, in an exact quantum treatment, the eigen-areas may receive 
quantum corrections, but this will 
{\it not} affect the exponential behavior. 

We now determine the expression for the 
outgoing modes which replaces eq. (\ref{Umodes}).
In this section, we shall not take into account the 
gravitational deformation induced by these modes.
We shall also neglect the effects that the
fluctuations of the geometry\cite{SORKIN, us} 
might have.
 The validity of both hypothesis will be discussed
after we have presented the mathematical outcome.
In the absence of back-reaction and metric fluctuations, the modes 
must be such that they correspond to excitations
of the vacuum state defined near the horizon. From 
the definition of $\Theta$, 
its relationship to the inertial light like 
coordinates $U_K, V_K$ (such that $ds^2 = 
- dU_K dV_K + r^2 d^2\Omega$ near the horizon)
 is of the form $dU_K = C e^{- \Theta} d\Theta $ at fixed $r$. 
This universal relation exhibits the exponential
Doppler shift which is the hallmark of horizons (except for 
extremal black holes). 
Therefore, we make the hypothesis that the new expression is 
\begin{eqnarray}
\phi_{\omega}(\Theta, R ) = D e^{i\omega U_K} = 
D e^{iC \omega e^{-\Theta}}
\label{photon}
\end{eqnarray}
in place of $e^{iC\omega e^{-\kappa t}} $,
see eqs. (\ref{gtoe}, \ref{etog}).
$D$ is a constant which plays no role.

We now assume that, to first order in $\gamma$ and 
as in eqs. (\ref{gtoe}, \ref{etog}),
the transition amplitudes are given by the 
``time'' integral of the product of the three waves.
Up to the {\it same} overall constant (see the discussion after 
eq. (\ref{Umodes})), 
they then are given by
\begin{eqnarray}
{\cal{B}}_{g \to e + \omega} &=& 
\int\! d\Theta\ \Psi_{BH + g}\;
\Psi_{BH + e}^*\; \phi_{\omega}^*
\quad , \nonumber\\
{\cal{B}}_{e \to g + \omega} &=& 
\int\! d \Theta\ \Psi_{BH +g}^*\;
\Psi_{BH +e}\; \phi_{\omega}^*
\quad . \label{exp}\end{eqnarray}

Using  eq. (\ref{PsiBHd}) and  eq. (\ref{photon}) and  by replacing
$\Theta$ by $\Theta + i \pi$ in either amplitude, one obtains
\begin{equation}
{ | {\cal{B}}_{g \to e + \omega}|^2 \over |{\cal{B}}_{e \to g + \omega}|^2 }
= e^{ (A_e - A_g)/4} \quad .
\label{ratio}
\end{equation}
This establishes that $(A_e - A_g)/4$, 
the difference of the horizon
areas if the detector is excited or not,
governs the equilibrium distribution of the detector's states.
The new distribution clearly corresponds 
to a micro-canonical distribution 
since we are considering exchanges
of energy between the black hole and the detector with no 
ADM mass change at spatial infinity. This confirms the interpretation
of $A/4$ 
as the statistical entropy of the black hole 
since $e^{A/4}$ enters in eq. (\ref{ratio}) as the quantum
degeneracy of the initial and final black hole states.
In this we confirm what has been found in \cite{HHR,suh2,KK}.

Eq. (\ref{ratio}) replaces the canonical expression of 
eq. (\ref{Prob})
which is governed by the energy change $E_e - E_g$ and 
by Hawking temperature $\kappa / 2 \pi$. 
To first order in $E_e - E_g$, energy conservation 
and the (static version\cite{WALDBOOK} of the) 
first law (i.e. $dE_{detect.} = - dM_{hole} = - \kappa dA/8 \pi $)
guarantee that the new expression 
gives back Hawking's result eq. (\ref{Prob}).
The correction to this first order approximation is governed by the 
specific heat of the black hole. 
Thus it is completely negligible for large black holes. 
Therefore the main 
changes from eq. (\ref{Prob}) to eq. (\ref{ratio})
are conceptual.
First, energy conservation is now built in 
through the use of the extremised total action.
Secondly, the thermalization of the detector no longer 
reveals that a thermal flux of photons is emitted by the
hole but more fundamentally that the detector is in contact 
with a reservoir whose entropy is ${A/4}$.
Thirdly, the geometry jumps from one stationary
situation to another one without smooth (classical)
interpolation between them.

\vskip .3 truecm   
{\bf Discussion}

\noindent
In order to reveal the origin of these qualitative changes
and to justify the hypothesis we made,
it is appropriate to rewrite the transition amplitudes in
terms of  the complete system:
black hole, detector and radiation field.

To quantize the whole system requires to consider the
 Wheeler-DeWitt constraints
\begin{equation}
H^{BH+detect. + \Phi}_\mu |{\Psi}\rangle =0
\quad . \label{const}\end{equation}
When $t$, the lapse of proper time at spatial flat infinity is fixed,
 the Wheeler-DeWitt equation is supplemented\cite{CT} by 
the following equation:
\be
i \partial_t |{\Psi}\rangle = \hat M_{ADM} |{\Psi}\rangle
\ee
Starting from the WDW equation, 
this shows that one recuperates
the notion of a Schr\"odingerian
evolution in terms of the (coordinate-invariant) time $t_\infty$
when one imposes that the 3-geometries are asymptotically flat.
Note also that this equation is 
the quantum version of eq. (\ref{deltaS'}), thereby justifying the
classical analysis of that Section.

Similarly, in the presence of an inner boundary,
when working at fixed opening time $\Theta$,
the Wheeler-DeWitt equation is supplemented by
another remarkable equation\cite{CT}:
\be
i \partial_\Theta |{\Psi}\rangle = - {\hat A \over 8 \pi } |{\Psi}\rangle
\quad . \label{CarT}
\ee 
Notice the different signs of $\hat A$ and $\hat M$ in these two
equations. This is because $\hat A$ comes from an inner boundary and
$\hat M$ from an outer one. This flip of sign can already be seen at 
the classical 
level by comparing eqs. (\ref{deltaSHam}) and (\ref{deltaSHam2}).
Indeed, eq. (\ref{CarT}) corresponds to the quantized version
of the Hamilton-Jacobi equation $\partial_\Theta S = A$ 
which follows from eq. (\ref{dddd}).
In eq. (\ref{CarT}), the operator 
$\hat A$ is defined from the behavior of the 
3-geometries as one reaches the horizon. When the local WDW constraints
are satisfied, $\hat A$ parametrically depends on the
matter and gravitational configurations from the horizon
till $r= \infty$. (A simple example of this dependence is provided by
the solutions (\ref{PsiBHd}) of equation (\ref{CarT}) where the
eigenvalue $A$ depends on the state of the detector.)

In our derivation, we first assume that, in the absence of interactions
governed by $\gamma$,
the energy of the radiation field (whether or not it is 
in an excited state) does  not influence the geometry. 
This amounts to assume that the wave functions
of the initial and final free states factorize:
\begin{eqnarray}
|\Psi_{in}(\Theta)\rangle
&= & 
|\Psi_{BH + g}(\Theta )\rangle
\otimes |0_U(\Theta )\rangle
\quad , \nonumber\\
|\Psi_{fin} (\Theta)\rangle& =& 
|\Psi_{BH + e }(\Theta)\rangle 
\otimes |\Psi_{\Phi_\omega }(\Theta)\rangle 
\quad .
\end{eqnarray}
This factorization is equivalent
to postulate in a path integral formulation that the 
total action
splits as a sum: $S_{BH + det + \Phi} = S_{BH + det} + S_\Phi$ wherein the 
latter is evaluated in the background defined by $S_{BH + det}$.
As shown in \cite{clbfa}, this approximation means that the 
recoil of gravity due to the energy of the radiation field has been 
taken to account to first order only. The 
corrections to this linear approximation are governed by
gravitational interactions of the form
$\int \!\! \int T G T$ where $T$ is the
energy momentum tensor of the radiation field and $G$ 
the Green function of the background degrees of freedom\cite{bk}. 
These interactions contain three parts. The
first describes the gravitational interactions 
of the photon and the detector. Neglecting these 
is probably legitimate if the detector is far from the black hole. 
The second part concerns the gravitational self interactions of the
photon. These will be studied in the context of the shell model of KKW
in the next section and will be shown
to confirm eq. (\ref{photon}). 
The third part describes the interactions among the 
quanta present in Unruh vacuum. In a dilute gas
approximation these are neglected.

In brief, under the assumption of factorizable wave functions,
the time dependence of the black hole + detector 
waves are given by eq. (\ref{PsiBHd}) in virtue of 
eq. (\ref{CarT}). And eq. (\ref{photon}) follows from usual 
Unruh boundary condition.
This last assumption also means that we postulate that the
near horizon fluctuations have no significant effect on 
Unruh vacuum. One might indeed fear that 
these fluctuations\cite{SORKIN,us} would destroy its characterization.
We are tempted to believe however that this is not the case.
First the recent body of works initiated by Unruh
on acoustic black holes\cite{dumbUnruh,dumbus} 
pleads in favor of this belief. In these
works indeed, it was shown that there is an adiabatic decoupling between
the low energy physics governed by the surface gravity
and the high energy regime. This explains why modifications
 of the dispersion relation at high frequencies affect neither
 the low energy properties of Hawking radiation nor the 
characterization of Unruh vacuum. That this
also applies to the case of near horizon fluctuations 
is the subject of current work\cite{FP}.
 
The second hypothesis concerns the transition amplitudes.
To obtain them, one should incorporate
the interaction Hamiltonian, eq. (\ref{hinter}), 
in the Wheeler-DeWitt constraint, 
represented here by eqs. (\ref{const}) and (\ref{CarT}).
To first order in $\gamma$, the modified propagation 
is expressed in terms of matrix elements of this Hamiltonian 
sandwiched by the free wave functions, as usual
for first order in the Born series. 
This is also true when working with the solutions of the 
Wheeler-DeWitt equation\footnote{
In fact the derivation of eq. (\ref{sssu}) closely follows that
of transition amplitudes in quantum cosmology
as performed in \cite{wdwpt}.
In both cases the wave functions appearing in matrix elements 
are WKB solutions of the Wheeler-DeWitt equation
which govern free propagation.
Another useful analogy is provided 
by the Unruh effect\cite{U}.
In that case, the given trajectory (which plays the role of  the 
classical geometry in black hole physics)
is replaced by WKB waves
in order to take into account recoil effects\cite{bfa}.}.
Thus to first order in $\gamma$ and up to 
an overall factor, the transition amplitude is given by
\begin{eqnarray}
{\cal{B}}_{g\to e + \omega}
&=& - i \gamma
\int d \Theta \ 
\langle \Psi_{BH+e} | \langle \Psi_{\Phi_\omega} | 
\left[ \hat \Phi( R ,\Theta) |e\rangle \langle d 
| \right] | \Psi_{BH + g }\rangle |0_U\rangle
\nonumber\\
&\simeq& -i\gamma \int d\Theta \  e^{-i ( S_{BH + e} -S_{BH + g })}
\times e^{-i S_{\omega}}
\label{sssu}
\quad .
\end{eqnarray}
In the second line, we have written the phase factors in terms 
of the action of the black hole + detector system and that of 
the radiation field.
This is to emphasize that only differences of actions appear in the
integrand. Indeed, in the first factor, the action of the black hole 
alone cancels
and one is left with the difference 
due to the change in the detector's state.
Similarly, in the second factor, $S_{\omega}$ is the change in action
of the radiation field due to the insertion of the field operator 
at $R, \Theta$. Using the stationary character of the states for
the first factor and Unruh boundary condition for the second
 lead to eq. (\ref{exp}) and hence to our central result eq. (\ref{ratio}).

To obtain further insight about eq. (\ref{sssu}), it is interesting to
show how it gives back the conventional amplitude
$B_{g\to e + \omega}$ obtained in a given background,
see eq. (\ref{gtoe}).
To recover this expression, it suffices to evaluate the difference in actions
appearing in eq. (\ref{sssu}) to first order in the energy change:
\begin{eqnarray}
S^{BH + det. + \Phi}_{fin}(\Theta) - 
S^{BH + det. + \Phi}_{in}(\Theta)  &=&
 \omega  \partial_\omega S_{\omega}(\Theta) + (E_e - E_g)
\partial_{E_{det.}} S_{BH+det.}(\Theta) 
\nonumber\\ &=&  \omega U_K(\Theta) +(E_e - E_g)t(\Theta)
\quad .
\end{eqnarray}
In deriving this, we have first used the splitting of the total 
action discussed above and then Hamilton-Jacobi 
equations. Indeed,   
$U_K$ is conjugate to the frequency $\omega$ 
and $t$ to $E_{detector}$.
It should be stressed that this recovery of 
the background field phases in the
limit of small energy differences is a generic 
feature\cite{bfa,clbfa,suh2}:
Whenever one takes into account a neglected heavy degree of 
freedom (here gravity), describes it by WKB waves and expands the 
resulting expressions for light transitions 
to first order in the light change, one
recovers the usual background field expressions in which
the heavy variable is treated classically.

Moreover this first order expansion ``commutes''
with the integration over $\Theta$. Thus it could equally be carried
out {\it after} having performed the $\Theta$ integration. In this second
form, one is expanding the exponential governing
transition rates, eq. (\ref{ratio}), to first order in $E_e -E_g$.
In this one recovers the first law of black hole mechanics.
This shows that the first law is nothing but the
Hamilton-Jacobi equation applied to the Euclidean sector 
in the absence of conical singularity:
\be
\partial_{E_{det.}} S^{Euclid.}_{BH+det.} = \partial_{E_{det.}}
(A/4)\vert_{M_{ADM}} = - 2 \pi / \kappa \, .
\ee

In view of the generic character of these features,
it is clear that our analysis also applies to charged or rotating
holes (see however the remark made after eq. (\ref{Umodes}))
and to cosmological or acceleration horizons.
Indeed all that is required is the evaluation of the 
phase factors entering eq. (\ref{sssu}). In this expression,
the second factor $e^{i S_\omega }$ presents no 
difficulty: it always encodes vacuum conditions
as one crosses the future horizon.
The first factor is more delicate since it requires
to solve Einstein's equations driven by the
energy of the detector.  However in the case 
of static Rindler like situations, the on shell action
gets its contribution only from the surface term
at the horizon. Moreover, since only differences
appear in transition amplitudes, eq. (\ref{sssu}) 
also applies to the accelerated cases. Indeed even though the
area of  acceleration horizon might be infinite, 
the change in the horizon geometry 
due to a finite change in the matter energy distribution 
is finite and well defined, see \cite{HHR}
for an explicit computation. 
In particular, it is local in the transverse directions $x^i_\perp$
when the change in the matter energy distribution 
is localized. This leads to finite and well defined
changes in on-shell actions for the gravity--detector system 
which furthermore gives back the conventional
background field result when linearized in $\Delta E_{detect.}$.

\section{The KKW model}\label{secKKW}

In the previous  calculation the change in area was due to the change
of the detector state. However the existence of a detector
is not intrinsic to black hole radiation: the detector was only used
to reveal the existence of the quanta of the radiation field.
Therefore we seek for an intrinsic derivation of 
black hole radiance in which the change in area is
due to the emission process itself.
In this description, the change in area
plays the role of the energy of the 
emitted quantum in Hawking's derivation.
To this end, we must introduce a model for the emitted quanta
which takes into account the deformation of the gravitational 
background. 
The simplest model is that of  KKW\cite{KW1,KW2,KK} in which 
one makes the hypothesis
that in the semi-classical limit and at high frequency the particles 
are described by self gravitating spherically symmetric light-like shells. 

The starting point of the KKW model is the 
situation analyzed in \cite{israel}. It describes the entire spherically
symmetric space time, solution of Einstein's equations,
 which results from the propagation of a light like shall. 
By Birkoff's theorem, both
outside and inside the shell the geometry is Schwarzschild.  
As in section \ref{SProb} we shall take the outside mass $M_{ADM}$
to be fixed whereas inside the residual mass $M(\lambda ) = 
M_{ADM} - \lambda$ depends on  $\lambda$,
the energy of the emitted shell
measured at $r=\infty$. 
In both geometries the shell follows an outgoing light like geodesic,
see \cite{israel}.
In what follows we shall use only the inner metric to describe the
trajectory, the action and the wave function of the shell. 
This choice will be justified after having presented
the results.

Inside the shell, in Eddington-Finkelstein coordinates the metric is
\begin{eqnarray}
ds^2= -(1 - {2 M(\lambda )  \over r})dv^2 + 2 dv dr + r^2 d\Omega^2
\end{eqnarray} 
and the trajectory of the shell satisfies
\begin{eqnarray}
dv &=& 2 {dr \over 1 - 2 M(\lambda ) / r_{sh}}
\quad ,\nonumber\\
v(r_{sh}) &=& 2 r_{sh} +  4 M(\lambda )  \ln (r_{sh} - 2M(\lambda )
 )\quad .\label{t2}
\end{eqnarray} 

As in the previous section it is a appropriate to introduce a
dimensionless time parameter defined near the horizon. The light like
version of $\Theta$ is
\begin{eqnarray}
V=\kappa(\lambda )v = \Theta + \kappa(\lambda )r + {1 \over 2} 
\ln (r-r_H(\lambda ))
\end{eqnarray} 
where $\kappa(\lambda )= {1 \over 4M(\lambda ) }$ is the surface gravity and 
$r_H(\lambda )= 2M(\lambda ) $ is the final radius of the horizon.
We shall also use as energy variable the area of the horizon
$A= \pi r_H^2(\lambda )$ 
rather then $\lambda $ because $A/8 \pi$ is the conjugate to $\Theta$
and to $V$. In
terms of these new variables the trajectory is
\begin{eqnarray}
V(r_{sh}; A)&=& \ln (r-r_H(A)) +2\kappa(A)r
\simeq \ln (r-r_H(A)) + O(r-r_H(A))\quad .
\label{Veq}
\end{eqnarray} 
Close to the horizon the log dominates and the
trajectory is expressed only in terms of
quantities locally defined. Therefore this expression 
characterizes radial
trajectories near all event horizons. As an illustration of this
universality one easily verifies that
for a charged non
extremal black hole, the trajectory near the horizon also takes the
form $V \simeq \ln (r-r_H)$ wherein the surface gravity 
does not appear. As in the former Section, we make the hypothesis that
the near horizon fluctuations do not alter the analytical behavior 
of this trajectory.

In order to obtain the modes characterizing the quantum propagation
of this shell we need its action. To obtain the action one could start
from scratch, that is from the Einstein-Hilbert plus matter
action and extremise it. This is the path followed in
\cite{KW1,KW2}. However, since we know the trajectories we can use a short
cut to obtain the Hamilton-Jacobi action, see also \cite{KK}.

Since the inside geometry is static, $A$ is a constant of motion. Therefore
the action can be written as
\begin{equation}
S(r, V; A) = {A V \over 8 \pi} + f(r,A)
\end{equation}
where $f$ is the Maupertuis action, that is $\partial_r f= p$ is the
momentum of the shell.
The classical trajectory follows from stationarity of $S$ with respect
to $A$, $\partial_A S = 0$. In the present case it implies
that $\partial_A f =- {V(r;A) / 8 \pi}$
where $V$ is given by eq. (\ref{Veq}).
After integration from $A$ to $A_0$, one obtains
\begin{equation}
S(r, V; A) ={(A - A_0) V \over 8 \pi}+
 \int^{A_0}_{A} 
{d \tilde A\over 8 \pi } \left( \ln (r - r_H(\tilde A)) + 
{\kappa(\tilde A) r } \right)
- g(r_0) \quad .
\label{SS}
\end{equation}
$A_0$ is the 
area of the horizon in the absence of shell, equal to $A_0 = 4 \pi
M^2_{ADM}$. 
With this choice, 
in the absence of the shell the action vanishes. This guarantees that
to first order in the shell energy $\lambda$ one identically recovers 
the action of a massless particle in the background geometry (at this
order it can be taken to be either the inner or outer geometry).
Notice also that $A_0$ is larger then $A$ since the area of the horizon
decreases when a particle of positive energy is emitted. 

We have also added an $r$ and $V$ independent integration constant
\be
g(r_0) = 
 \int^{A_0}_{A} 
{d \tilde A\over 8 \pi } \left( \ln (r_0 - r_H(\tilde A)) + 
{\kappa(\tilde A) r_0 } \right) \quad .
\ee
At fixed $A$ this function enables 
the initial momentum $p_0$ to be fixed arbitrarily 
and determine implicitly the initial radius $r_0$ or conversely to 
fix $r_0$ and determine implicitly $p_0$.
Technically this follows from $\partial_{r_0} S = -p_0(A,r_0)$.

The action eq. (\ref{SS}) is well defined on either side of the
smeared horizon, that is for both $r$ and $r_0$ greater than $r_H(A_0)$ 
or both less than $r_H(A)$. 
In each region, we can use it to
define (in the WKB approximation) the wave function of self
gravitating shells. For $r$ and $r_0$ greater than $r_H(A_0)$,
one has
\begin{equation}
\varphi_{\Delta A, +} =
e^{i S(V,r; \Delta A)}
\; \theta (r- r_H(A_0))
\quad . \end{equation}
This describes the classically allowed propagation of a shell outside
the horizon. Similarly we can define a wave function 
\be
\varphi_{\Delta A, -}^* = e^{i S(V,r; \Delta A)} 
\; \theta (r_H(A)-r)
\label{cccc}
\ee
living only in the region $r, r_0 < r_H(A)$. 
These definitions are in strict analogy with
the wave functions $\varphi_{\lambda \pm}$ defined in
eq. (\ref{outmodes}) and reduce to them in the limit $\lambda \to
0$. In particular the complex conjugation in eq. (\ref{cccc}) arises
because behind the horizon $\Theta$ time runs backwards.
Notice that we did not write the prefactors of these waves.
To conform ourselves to second quantized rules, we should have
introduced prefactors such that the Wronskian be unity.
We shall not pursue this since the prefactors play no role
in what follows, i.e. the determination of the pair creation probability.

Between
$r_H(A_0)$ and $r_H(A)$ one needs a prescription to define the
logarithm. As explained at the end of section 2, 
the definition of the action can be used to encode the Unruh boundary
condition. 
The analytical specification imposes that the $\phi_{\Delta A, +} $ mode,
the analog of $\phi_{\lambda, +} $ of eq. (\ref{US}), 
\begin{equation}
\phi_{\Delta A, +} =
e^{i S_U(V,r; \Delta A)}
\label{phiDA}
\end{equation}
be analytical and bounded in the upper half of the complex $r$ plane
at fixed $V$. This leads to the globally defined action
\begin{eqnarray}
S_U(r, V; A, r_0 > r_H(A_0)) ={(A - A_0) V \over 8 \pi}
\quad\quad\quad\quad\quad\quad\quad\quad\quad\quad
& &\nonumber\\
+ \int^{A_0}_{A} 
{d \tilde A\over 8 \pi } \left[ \ln | r - r_{H}(\tilde A) | + i\pi
\theta(r_H(\tilde A) - r) + O(r - r_{H}(\tilde A)) \right]
- g(r_0)\quad .& &
\label{SS2}
\end{eqnarray}
Thus as $r$ goes from $r>r_H(A_0)$ to
$r<r_H(A)$ the action acquires an imaginary
part equal to  
\be
Im S_U = {A_0 - A \over 8}
\label{imsu} \quad .
\ee 
This simple result is due to the fact that the log in eq. (\ref{SS})
comes with an $\tilde A$ independent weight. 
As emphasized after
eq. (\ref{Veq}) the origin of this independence  follows 
from the universal form
of light like outgoing trajectories near a future 
horizon: $ V = \ln (r -r_H(A))$. 
Therefore eq. (\ref{imsu})
universally follows from this behavior and 
from Unruh's prescription for the analytical behavior of
the modes. This is the essential kinematical result of this section.

Using this result, we can write $\phi_{\Delta A, +}$ as the linear combination
\be
\phi_{\Delta A, +} = 
\varphi_{\Delta A, +} +e^{-\Delta A / 8} \varphi_{\Delta A, -}^*\quad . 
\ee
wherein only the relative weight of $\varphi_{\Delta A, +}$
and $ \varphi_{\Delta A, -}^*$ has meaning, c.f. the above discussion
about the normalization of the Wronskian.
In section \ref{two}, the technique of Bogoljubov transformation
enabled us to identify the square of the ratio of the weights of $
\varphi_{ +}$ and $\varphi_{ -}^*$ with the probability to emit a
pair, see eq. (\ref{eqex}). In the presence of backreaction,
the same relation still holds for rare and energetic events.
Thus we obtain
\be
P_{\Delta A} = e^{- 2 Im S_U} =  e^{- \Delta A /4}
\label{grres}
\ee
in place of eq. (\ref{eqex})
and in agreement with eq. (\ref{ratio}) and \cite{KK}. 
This result can probably be generalized to rotating holes
and to cosmological or acceleration horizons. But this will require that
one refines the procedure so as to take into account
the transversal directions. We hope to return to this 
problem.

It should be stressed that the 
justification of the identification of $e^{- 2 Im S_U}$ 
as the transition probability 
is more delicate in  the presence of backreaction 
than in the free field theory. This point is discussed 
below. 

\vskip .3 truecm
{\bf Discussions}

\noindent
To prepare the discussion, it is appropriate to
compare our treatment with the original derivation of
KKW\cite{KW1,KW2,KK}.
A first difference with the calculation of KKW is that 
we use $A, V$ instead of $\lambda,t$ as energy and time
variables. The passage from one to the other is straightforwardly
implemented by using the $\lambda$-dependent 
Jacobian $dA/d\lambda$ in the action. The advantage of
using the $A, V$ variables from the start is that it is then manifest
that the result for the probability of emission $P = e^{-\Delta A/4}$
is no accident, but follows from the universal form of classical
trajectories near a horizon. 

A second difference is that 
KKW work with fixed black hole mass $M$ and with a varying
ADM mass $M+ \lambda$. In an empty geometry,
this can be shown to be mathematically
equivalent to working as we do with a varying black hole area
and fixed ADM mass. However as discussed in section \ref{boundary} 
the second description reflects better the physics of the 
emission process wherein the black hole
loses energy to the radiation while the mass at infinity stays constant.
Then, when working at fixed ADM mass, the time at infinity cannot be
used to parameterize the propagation of the shell, see \cite{PW}
for a general proof of this super-selection rule. Thus, as in Section 4,
it is through the $\Theta$ (or $V$) dependence 
that one recovers the notion of evolution.
For non empty geometries, we conjecture that it will be mandatory
to work with a varying horizon area, i.e. to use the variables $A, V$
to parameterize the process.

Thirdly we have implemented the Unruh boundary condition on
the area eigenstates $\phi_{\Delta A, \pm}$ through
the analytical behavior in $r$ as it goes from one side of the 
horizon to the other. 
In \cite{KK} instead, the Unruh boundary
condition is implemented by fixing the initial momentum
$p_0= - \partial_{r_0} S$ at $V=0$. These two procedures are
equivalent. Indeed, in both cases, the definition of
$\ln(r-r_H)$ in the action is obtained by imposing positive frequency in
$i\partial_r$ across the horizon.

The boundary condition of fixed large positive momentum $p_0 \gg \lambda$
also shows the relation with the approach 
of section \ref{SProb}. Indeed the action of a self gravitating shell with
initial constant momentum $p_0$ can be shown\footnote{
This is obtained by replacing the equation for the trajectory near the 
horizon
$V = \ln { r - r_H(A) \over r_0 - r_H(A)}$ in the equation for the
momentum
$p = \int_A^{A_0} {d \tilde A \over 8 \pi } {1 \over r - r_H(\tilde
A)}$ 
to obtain
$p = p_0 e^{-V}$. Integrating the 
Hamilton-Jacobi equation $\partial_r S = p$,
one finds $S_{p_0} = p_0 r e^{-V} + F(p_0,V)$ where $F$ is an $r$
independent integration constant. Then imposing that on
the equations of motion the
action be stationary with respect to variations of $p_0$, one finds 
eq. (\ref{actionp0}) with $\partial_{p_0} h(p_0) = r_H(r_0, p_0)$.}
to be
\be
S(r,V)_{p_0}=[ p_0 r + h(p_0)] e^{-V}
\quad . \label{actionp0}\ee
The assumption of KKW that in the presence of backreaction modes have
the form $\phi = e^{i S_{p_0}}$ is therefore equivalent to the
assumption made in section \ref{SProb}, 
see eq. (\ref{photon}), that the wave function of
particles in Unruh vacuum is proportional to
$e^{iC \omega e^{-\Theta}}$ at constant $r$.

We now turn to the delicate question of identifying twice the 
imaginary part of the action 
to go from one side of the horizon to the other with the (log of the)
probability to emit a pair. 
In addressing this question one faces a double problem. First,
due to gravity,
the emission process is no longer linear. 
Indeed when two shells of energy $\lambda$ are emitted,
their action is not twice that of one particle.
Therefore,  probabilities for multi-particle production 
are no longer obtainable from those for single particle emission as
they are in a linear field theory.
This means that the machinery of Bogoljubov transformations
no longer applies. The second problem is due to the use of 
WKB approximations for the wavefunctions which 
are given in terms of the action of a {\em single} shell. Because of
this we 
are certainly not in a position to describe multi-particle effects,
that is higher order effects in the tunneling amplitude $e^{-Im S_U}$.

We shall now sketch how one can deduce, to 
leading order in $e^{-Im S_U}$,
the probability of particle production from the properties
of the wave functions, without resorting to Bogoljubov transformations. 
To present our method, we
first return to the analysis in the absence of
backreaction and consider the following matrix element 
\ba
{\langle 0_{U}| \int d\tilde v e^{i \lambda \tilde v}
\Phi(R=+\infty,\tilde v) 
\Phi(r,v)  |0_U\rangle
 \over \alpha^2_\lambda }
&=&
{\langle 0_{U}|  a_{out,\lambda +}
\Phi(r,v) 
 |0_U\rangle
 \over \alpha^2_\lambda }
\nonumber\\
&=&{ \phi_{\lambda +}^*(r,v) \over \alpha_\lambda} = 
\varphi_{\lambda +}^*(r,v)
+ {\beta_{\lambda} \over  \alpha_{\lambda}} \varphi_{\lambda -}(r,v)
\label{phi*}  \quad . 
\label{rat4}
\ea
When
$r>2M$, it has a simple interpretation. 
It is the amplitude for an out-particle created at $(r,v)$ to be
found at ${\cal I}^+$ with energy $\lambda$. This is a classical
process governed by the action $ \int_{r}^{+\infty} dr'
p_\lambda (r')$ where $p_\lambda = 2 \lambda/( 1 - 2M /r)$ is the classical
momentum of the particle.  The normalization 
has been chosen so as to describe one out-particle,
i.e. to have a unit current (Wronskian) for $r>2M$.
In a linear field theory this  normalization can be calculated exactly 
and is given by the 
factors of $\alpha_\lambda$, see eq. (\ref{US}). 
It is thus proportional to $1 + O(e^{-2\pi \lambda/\kappa})$.
For self-gravitating shells, this
normalization might be different
 but will remain proportional to $1 + O(e^{-2Im S_U})$.

When $r<2M$,  this matrix element has also a simple interpretation 
because $\Phi(r,v)$ and $a_{out,\lambda +}$ commute. 
It defines the amplitude for a pair to be 
emitted by the black hole. Indeed, it
defines the transition amplitude from $ |0_U\rangle$ to the state
$\Phi(r,v) a^\dagger_{out,\lambda +}    |0_U\rangle /\alpha^2_\lambda$. 
One member of the pair is outside the
horizon and has energy $\lambda$, the other is behind the 
horizon at $(r,v)$. 
>From this we deduce that 
 the probability $P_\lambda$ to create a pair is given by
the current $\beta^2_\lambda / \alpha^2_\lambda $ carried by 
$\phi_{\lambda +}^*/ \alpha_\lambda $ for $r<2M$.

This expression for $P_\lambda$ can in turn be used to show how 
$P_\lambda$ can be expressed in terms of the classical action.
Recall that the prescription for
defining the Unruh mode $\phi_{\lambda +}$ for $r<2M$
is such that its phase, i.e. the action $\int p_\lambda dr$, is
obtained by analytically continuing $r-2M$ in
the upper half complex plane at fixed $v$. This fixes
the ratio of the amplitudes $\phi_{\lambda +}$ on each
side of the horizon. Thus the probability $P_\lambda$
is given by                   
\be
P_\lambda = {
|\phi_{\lambda,+}(r>2M,v)|^2 \over |\phi_{\lambda,+}(r<2M,v)|^2}
= e^{-2 Im
\int_{r<2M}^{r>2M}dr' p_\lambda (r')} 
= e^{- 2\pi \lambda/\kappa}
\label{nR}
\ee
Note that both the normalization $\alpha_\lambda$ 
and the relativistic prefactors $(4 \pi \lambda)^{-1/2}$ cancel in this ratio.
This follows from the fact that $P_\lambda$ is given
by the ratio of the currents carried by $\phi_{\lambda,+}$ on each side
of the future horizon.

Upon taking into account the gravitational backreaction, 
the above argument still applies if two conditions are met.
First the WKB approximation of the wave functions must be
valid. This guarantees that we can use the classical action to evaluate
the relative amplitude of the Unruh modes on each side of the horizon.
Secondly, the wave function of the total system, black
hole + radiation field, must be (approximatively) 
factorisable\footnote{When 
the shell is very close to the horizon, 
this factorization probably breaks down
because of its momentum $\partial_r S$ 
($=p_\lambda (r)$ in a given background) 
is arbitrary large (trans-Planckian) since it diverges. 
In the presence of backreaction, 
the new momentum $\partial_r S_U$ diverges logarithmicly
as $r \to r_H(A)$, see eq. (\ref{SS2}).
However, once $\partial_r S_U$ is much smaller than the Planck mass,
the factorization should apply. The crucial point is the 
existence of an {\it intermediate} region in which the
log is dominating the action $S_U$, this requires $\partial_r S_U
\gg \lambda$, and in which the factorization applies, this 
requires $\partial_r S_U \ll M_{Planck}$. In this region, the 
usual analytical characterization of Unruh vacuum still applies\cite{Teddy}
and this is {\it sufficient} to obtain eq. (\ref{grres}).
It is interesting to notice that the same logarithmic behavior
(in an {\it intermediate} region) also explains
the absence of modifications to Hawking radiation when one 
mutilates the dispersion relation\cite{dumbUnruh, dumbus}} 
into the wave function of the shell times the rest. 
Both require that  the energy of the shell be large, 
i.e. $\lambda /\kappa \gg 1$. 

In quantum mechanics such a factorization is a good
approximation whenever there is a hierarchy in 
the degrees of freedom: very heavy
degrees of freedom that can be treated in the WKB approximation (the
nuclear degrees of freedom in the case of a Rydberg electron orbiting
an atom or molecule, the radius of the
universe in quantum cosmology, the black hole in the present case);
 moderately energetic degrees of freedom 
that can also be treated in the WKB
approximation (the Rydberg electron, heavy matter in cosmology,
rare energetic particles emitted by the black hole); and degrees of
freedom that must be treated quantum mechanically (the inner electrons,
the other matter degrees of freedom in cosmology, less energetic
particles emitted by the black hole). 
The very heavy and the moderately 
heavy degrees of freedom 
propagate semi-classically in the mean potential due to the other degrees of
freedom (for a detailed treatment of the Born-Oppenheimer approximation
in a dynamical context, see \cite{DTK} for atomic physics and \cite{WDW}
for cosmology). Moreover, the probabilities of the rare quantum
transitions of the moderately energetic degrees of freedom can be obtained
from the sole properties of their wave function.
The reason is that 
the wave functions of both the heavier and the lighter degrees cancel
out from the bra-kets which represent transition amplitudes.
An illustration of such  cancellations is given in eq. (\ref{sssu})
where only  the difference in actions
due to the process itself enter in the transition amplitude.

In brief, when both conditions are met, 
the former analysis performed in the
linear theory applies. This guarantees that to first order in 
$e^{-Im S_U}$, eq. (\ref{grres}) is correct. This is also 
what has been adopted in \cite{KW1,KW2,KK}.

\section{Conclusion}

Let us further discuss the hypothesis we made.
First we
supposed that, in the presence of backreaction, physics in the
neighborhood of a black hole is described by 
the total action $S_{gravity + matter}$. 
Second we supposed that the wave function of the total system
 factorizes into a piece describing the emitted particle 
and the rest describing the black hole and the other
matter degrees of freedom. Finally we supposed that the wave functions
governing heavy degrees of freedom could be approximated 
by a WKB form. 

The first hypothesis may seem completely evident, 
but one must recall that there are scenarios, such 
as the brick wall model of 't Hooft, in which it is not necessarily
true. Furthermore a recent critique of the KKW model\cite{critique} 
does not recover eq. (\ref{R}) because its matter model does not derive
from an action principle. The validity of the last two hypothesis was
discussed at the end of the preceding section and in \cite{KW2}.

Upon making these hypothesis one finds that the probability of the
black hole of emitting a particle of energy $\lambda$ is given by
\begin{equation}
P_{M \to M - \lambda} = N(\lambda, M ) e^{- \Delta A (\lambda, M) /4
  } \ .
\label{R2}
\end{equation}
We have emphasized the universal validity of this result by
giving two different derivations. Its importance lies 
in two facts. First it provides statistical foundations to  
black hole thermodynamics since it applies to every quantum emission 
and since it is governed by the 
induced quantum change (i.e. the recoil) of the horizon area.
But, because we have not identified the microcanonical states, and
because we work in a WKB approximation, it is only valid in a
mesoscopic sense discussed below.
Secondly since the derivation of eq. (\ref{R2}) also 
applies to any quantum emission arising in the presence of an 
event horizon, this statistical interpretation is also valid
for all event horizons.

To appreciate the first fact, let us
recall that the first law\cite{WALDBOOK} of black hole thermodynamics 
is a purely classical result which is obtained 
by comparing two slightly {\em different} solutions 
of Eintein's equations. On the other hand Hawking's derivation 
of black hole radiation is obtained by quantizing 
a matter field in a {\em given} geometry. Thus, there is no 
a priori reason why these two results obtained in so different 
settings should be related or even compatible.
It turns out however that they are consistent with each other
and this is the basis for the successful thermodynamics of black 
holes.

The origin of this complementarity can be easily understood 
from the ``improved'' derivation of Hawking radiation
presented here. Since energy conservation is implemented from the start,
one takes into account the change of the black hole geometry 
due to the emission itself and one finds that the emission  probability
 is governed by a difference of matter+gravity 
actions $S_{m+g}(final) - S_{m+g}(initial)= \Delta A/4$. From 
this macrocanonical result
one recovers the canonical distribution found by Hawking by
expanding this difference to first order in the energy
change $\lambda$. 
In the same way the first law is also obtained in \cite{CT}
by this expansion.
It is therefore a mathematical necessity that they be consistent. 

The other important aspect of eq. (\ref{R2}) is that 
it indicates that 
black hole thermodynamics is not only valid in the mean 
(in the sense of a large ensemble of processes) but that it applies
to every quantum mechanical process.
Recall that
quantum mechanics predicts that ratio of the
decay rate to the absorption rate of
the same quantum is proportional
to the ratio of the number of final to initial micro--states
(at least when the interaction Hamiltonian can be treated as a
constant). When this is the case, 
eq. (\ref{R2}) leads to the identification
\be
\Delta A/4 = {\mbox{the change of the ln of the nbr. of horizon states}}\ .
\label{ent}
\ee
Moreover, since eq. (\ref{R2}) also applies to quantum emissions
near all event horizon, { \em eq. (\ref{ent}) also applies 
to all event horizons}.
The universality of this identification invites three
comments\footnote{As an additional point, recall that Hawking
radiation enlarged the domain of validity of 
the generalized second law to
processes where the black hole temperature can be treated as a
constant\cite{UW}. We conjecture that eqs. (\ref{R2}) and (\ref{ent})
can be used to enlarge its domain of validity 
to situations in which the canonical approximation breaks
down, both for black holes and other horizons.}.

Firstly, it leads to the notion of 
entropy density\cite{Teit3, C} per unit horizon area. Indeed, 
when considering local creation processes rather than 
global ones described by $s$-waves, the change in area
is local\cite{HHR} in the transverse directions $x^i_\perp$.
This  means that a finite and localized set of horizon states
are affected by the process.

Secondly, since eq. (\ref{ent}) applies to acceleration horizons,
i.e. to near vacuum configurations having entropy
measured by inertial means that can be very small, 
one is lead to conclude that
$A/4 $ determines the number of states {\it available} to the 
(accelerated) system in contact with the event horizon, i.e. with its heat. 
If this is the case, it implies that the black hole area 
should be viewed as a measure of the entropy accessible 
to the external world\cite{PP2,jac}. How to define its `actual' 
entropy and whether this latter will still be given by $A/4 $,
as is found for extremal black holes in string theory,
\cite{calmad} are important questions.

Thirdly, it should be stressed that 
the identification expressed by eq. (\ref{ent})
says nothing about the microscopic nature of the degrees
of freedom which constitute the states of event horizons entering in the
transition amplitudes we computed. All we can say is that when 
the three hypothesis we made are valid, the number of horizon states
{\it must} obey eq. (\ref{ent}). Therefore,  eq. (\ref{ent})
must be conceived as a universal {\it mesoscopic}\cite{vanK, SORKIN2} result,
somehow intermediate between the classical laws of thermodynamics
and the ultimate theory of quantum gravity.

\vskip 1. truecm

{\bf Acnowledgements}
The authors would like to thank Roberto Balbinot and Ted Jacobson for 
enjoyable discussions. They also thank the referee and
R. Brout for their criticisms which helped to clarify the manuscript.
S.M. is a ``chercheur qualifi\'e du FNRS''. He
would like to thank Utrecht university where part of this work
was carried out.

\end{document}